\documentclass[12pt]{article}
 \pdfoutput=1
 \usepackage{graphicx}
  \usepackage{epsfig}
  \usepackage{amsmath}
  \usepackage{amssymb}
  \usepackage{color}
  \newlength{\abstractwidth}
  \newcommand{\be}{\begin{equation}}
  \newcommand{\ee}{\end{equation}}
  \newcommand{\tr}{\text{tr}}
  
  \renewcommand{\title}[1]{\vbox{\center\bf{\Large{#1}}}\vspace{5mm}}
  \renewcommand{\author}[1]{\vbox{\center#1}\vspace{5mm}}
  \newcommand{\address}[1]{\vbox{\center\em#1}}
  \newcommand{\email}[1]{\vbox{\center\tt#1}\vspace{5mm}}
  \definecolor{darkgreen}{rgb}{0,.5,0}

\begin{document}

\begin{titlepage}
\rightline{MIT-CTP/4594}
\rightline{SU-ITP-14/20}
\begin{center}
\hfill \\
\hfill \\
\vskip 1cm

\title{Localized shocks}

\author{Daniel A. Roberts,${}^a$ Douglas Stanford,${}^{b,c}$ and Leonard Susskind${}^b$}

\address{$^{a}$ Center for Theoretical Physics {\it and} \\  Department of Physics, Massachusetts Institute of Technology \\
Cambridge, MA, USA  

\vspace{10pt}

$^{b}$ Stanford Institute for Theoretical Physics {\it and} \\ Department of Physics, Stanford University \\ Stanford, CA, USA 

\vspace{10pt}

$^{c}$ School of Natural Sciences, Institute for Advanced Study,\\ Princeton, NJ, USA}

\email{drob@mit.edu, stanford@ias.edu, susskind@stanford.edu}

\end{center}

  \begin{abstract}
We study products of precursors of spatially local operators, $W_{x_{n}}(t_{n}) \dots W_{x_1}(t_1)$,  where $W_x(t) = e^{-iHt} W_x e^{iHt}$. Using chaotic spin-chain numerics and gauge/gravity duality, we show that a single precursor fills a spatial region that grows linearly in $t$. In a lattice system, products of such operators can be represented using tensor networks. In gauge/gravity duality, they are related to Einstein-Rosen bridges supported by localized shock waves. We find a geometrical correspondence between these two descriptions, generalizing earlier work in the spatially homogeneous case.

  \end{abstract}
  \end{titlepage}

\tableofcontents

\baselineskip=17.63pt

\section{Introduction}
How is the region behind the horizon of a large AdS black hole described in the dual gauge theory? A number of probes of the interior have been proposed. These  include two-sided correlation functions \cite{Louko:2000tp,Kraus:2002iv,Fidkowski:2003nf}, mutual information and entropy \cite{Morrison:2012iz,Hartman:2013qma,Liu:2013iza}, and the pullback-pushforward \cite{Freivogel:2004rd,Heemskerk:2012mn} modification of the standard smearing function procedure \cite{Banks:1998dd,Balasubramanian:1998de,Bena:1999jv,Hamilton:2005ju}. These probes work well for young black holes, but for older black holes, or for black holes in a typical state,\footnote{Arguments have been made that black holes in a typical state do not have an interior geometry \cite{Almheiri:2012rt,Almheiri:2013hfa,Marolf:2013dba}. We will not address typical states in this paper.} they do not appear to be helpful. One way of stating the problem is as follows: the gauge theory scrambles and settles down to static equilibrium in a short time, during which two sided correlations decay and mutual information saturates. On the other hand, an appropriately defined interior geometry of the black hole continues to grow for a much longer time \cite{maldacenaTN}. What types of gauge theory variables can describe this growth?

Two possible answers have been suggested. Hartman and Maldacena \cite{Hartman:2013qma}  studied the time evolution of the thermofield double state and pointed out the relationship between the growth of the interior and the growth of a tensor network (TN) description of the state. Building on this work and ideas of Swingle \cite{Swingle:2009bg}, Maldacena \cite{maldacenaTN} suggested that the interior could be understood as a refined type of tensor network describing the state of the dual gauge theory. According to this picture, the overall length of the interior is proportional to size of the minimal tensor network that can represent the state.

A second suggestion focused on the evolution of the quantum state as modeled by a quantum circuit (QC). It was conjectured \cite{Susskind:2014rva} that the length of the black hole interior  at a given time is proportional to the computational complexity of the state at the same time. The computational complexity is the size of the minimial quantum circuit that can generate the state, and it is expected to increase linearly for a long time $\sim e^S$. In \cite{Stanford:2014jda}, two of us checked a refinement of this relationship for a family of states, corresponding to the spherical shock wave geometries constructed in \cite{Shenker:2013yza}.

There are strong reasons to think that the tensor network and quantum circuit descriptions are essentially the same thing. A QC is a special case of a TN. It has some special features, such as time-translation invariance and unitarity of the gates, which are not shared by the most general TN. But as noted in \cite{Hartman:2013qma}, these features are also necessary for a TN to be able to describe the black hole interior. Thus we will assume that that the TN and QC representations are the same.

In this paper, we will continue exploring the relationship between tensor network geometry and Einstein geometry. We will work in the setting of two-sided black holes. Our hypothesis, following \cite{maldacenaTN}, is that the geometry of the minimal tensor network describing the entangled state is a coarse-graining (on scale $\ell_{AdS}$) of the Einstein-Rosen bridge connecting the two sides. We will consider TN and Einstein geometry associated to products of localized precursor operators, each of the form
\be
W_x(t_w) = e^{-iHt_w}W_x e^{iHt_w}.\label{pre}
\ee
For $t_w = 0$ this precursor is simply $W_x$, an operator local on the thermal scale. But as $t_w$ advances, it becomes increasingly nonlocal. In a lattice system, such an operator can be represented in terms of tensor networks, and we will argue on general grounds that the characteristic TN geometry of a single precursor consists of two solid cones, glued together along their slanted faces. General products of precursor operators
\be
W_{x_n}(t_n)...W_{x_1}(t_1)
\ee
can also be represented in terms of tensor networks, with geometries that we will characterize in \S~\ref{1}.

Spatially homogeneous precursors were analyzed using gauge/gravity duality in \cite{Shenker:2013pqa,Shenker:2013yza,Leichenauer:2014nxa,Stanford:2014jda}. Their action on a thermal state corresponds to adding a small amount of energy to an AdS black hole. As $t_w$ is increases, the stress energy is boosted and a gravitational shock wave is produced. Products of precursors create an intersecting network of shock waves behind the horizon \cite{Shenker:2013yza}. In \S~\ref{2}, we will extend this analysis to the case of spatially localized precursor operators. We will study the spatial geometry of the two-sided black hole dual to $W_x(t_w)$, and find that it has the same ``glued cone'' geometry we inferred for the TN. More generally, we will see that the ERB geometry dual to multiple perturbations, local at different times and positions, agrees with the expected structure of the TN. More specifically, it agrees on scales large compared to $\ell_{AdS}$. This generalizes \cite{Stanford:2014jda} and provides a wide range of examples relating TN and Einstein geometry \cite{maldacenaTN}.

A central object in our analysis will be the size and shape of a precursor operator. In the spin chain and holographic systems that we study, precursors become space-filling, covering a region that increases outwards ballistically with respect to the time variable $t_w$. This behavior can be diagnosed using the thermal trace of the square of the commutator,
\be
C(t_w, |x-y|) = \tr\left\{\rho(\beta)[W_x(t_w),W_y]^\dagger[W_x(t_w),W_y]\right\} \label{thermal-commutator}
\ee where $W_x(t_w)$ is the precursor, and $W_y$ is a local operator at point $y$. For simplicity, let us consider unitary operators, so that the maximum of this quantity is two. We will define the size $s[W_x(t_w)]$ as the $(d-1)$ dimensional volume of the region in $y$ such that $C$ is greater than or equal to one.\footnote{It would be more precise to optimize over all operators $W_y$ at location $y$. However, for a suitably chaotic system, this step is not necessary: the butterfly effect suggests that any operator will do.} In the examples that we consider, this region consists of a ball centered at location $x$. We will define the radius of the operator $r[W_x(t_w)]$ as the radius of this ball.\footnote{\label{growthvsmovement}It is important to distinguish growth from movement: the growing operator is not a superposition of operators at different locations. If this were the case, the commutator $C$ might be nonzero in a large region, but it would be numerically small. The fact that the commutator is order one indicates that the operator $W_x(t_w)$ is sum of complicated products, each summand including nontrivial operators at most sites within the region of size $s[W_x(t_w)]$. This is the sense in which the operator is space-filling.}

We will see that the radius increases linearly with $t_w$. In the spin chain system, we will check this numerically. In the large $N$ holographic system, we will use the geometry of the localized shock wave to determine
\be
r[W_x(t_w)] \approx v_B(t_w - t_*).
\ee
Here $t_* = \frac{\beta}{2\pi}\log N^2$ is the scrambling time, and the ``butterfly effect speed'' $v_B$ is $\sqrt{\frac{d}{2(d-1)}}$ \cite{Shenker:2013pqa}, where $d$ is the spacetime dimension of the boundary theory. This expression is negative for times $t_w < t_*$, indicating that the $W$ perturbation has not had an order one effect on any local subsystem. However, after the scrambling time, the precursor grows outwards at speed $v_B$; this should be understood as the spread of the butterfly effect.

\begin{figure}[ht]
\begin{center}
\includegraphics[scale=1]{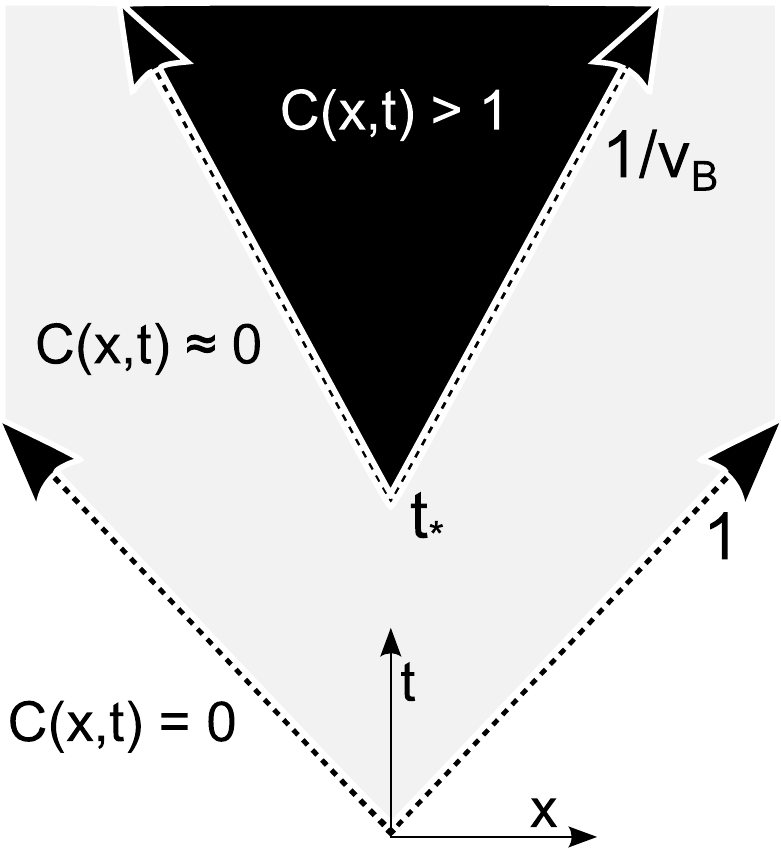}
\end{center}
\caption{A plot of $C(x,t)=\tr\left\{\rho(\beta)[W_x(t),W_0]^\dagger[W_x(t),W_0]\right\} /\, \tr\left\{\rho(\beta)W \, W \right\}^2$ illustrating the various ``cones'' related to causality and the butterfly effect. The causal light cone is defined by $t=x$. For $t<x$ the operators are spacelike separated and the commutator $C(x,t)$ must vanish by causality (white region). For $t<x/v_B + t_*$, $C(x,t)$ will be approximately zero (light gray region). The butterfly effect cone is defined by $t=x/v_B + t_*$ such that $C(x,t)=1$ on the butterfly effect cone, and the commutator will quickly obtain the value $C(x,t)=2$ on the inside (black region).}
\label{fig:light-cones}
\end{figure}

{\bf Added in [v3]:} The speed $v_B$ combined with the scrambling time $t_*$ can be thought of as defining a ``butterfly effect cone'' inside of which the commutator \eqref{thermal-commutator} is large. In Fig.~\ref{fig:light-cones} we sketch this region along with the standard light cone to emphasize the difference. The point is that even if operators are timelike separated with respect to the causal light cone, if they are spacelike separated with respect to the butterfly effect cone the early perturbation cannot yet have a nontrivial effect on the later operator. Therefore, it is the butterfly effect cone that controls the size of a precursor and the development of chaos.

\subsection{Some terminology}

For the convenience of the reader we will list some terminology used in this paper.

\begin{itemize}
\item $d$ refers to the space-time dimension of the boundary theory.
\item $t_*$ is the scrambling time $\frac{\beta}{2\pi}\log N^2$, where $N$ is the rank of the dual gauge theory.
\item A precursor $W(t_w)$ of an operator $W$ is given by $e^{-iHt_w}We^{iHt_w}$. A localized precursor $W_x(t_w)$ is a precursor of an approximately local operator $W_x$.
\item A localized precursor $W_x(t_w)$ is associated to a region of influence, i.e. the region in which local operators have an order-one commutator with $W_x(t_w)$. 
\item The $(d-1)$-volume of this region defines the size $s[W_x(t_w)]$. In a qubit model, the size indicates the number of qubits affected at $t = 0$ by the action of a single qubit $W_x$ a time $t_w$ in the past. 
\item The radius of the affected region is $r[W_x(t_w)]$. Size and radius are related: size is the $(d-1)$-volume of a ball of radius $r.$ We will see $r \approx v_B(t_w -t_*)$, where $v_B$ defines the speed at which the precursor grows.
\item  $\Sigma_{max}$ is the spatial slice of maximal $d$-volume that passes through the ERB, from time $t = 0$ on the left boundary, to $t = 0$ on the right. $\Sigma_{dec}$ is similarly defined, but it maximizes a functional obtained from the volume by dropping transverse gradients.
\end{itemize}

\section{Qubit systems}\label{1}
Black holes of radius $\ell_{AdS}$ have entropy of order $N^2$ where $N$ is the rank of the dual gauge group. This is the number of degrees of freedom of a single cutoff cell of the regulated gauge theory. We can think of such black holes as units out of which larger black holes are built. Technically this means that the gauge theory can be represented as a lattice of cells, and that in order to represent the thermal state at temperature $T$, the coordinate size of a cell  should be no bigger than $T^{-1}$.

A perturbation applied to a  degree of freedom on a specific site will evolve in two ways: it will spread out through the $N\times N $ matrix system through fast-scrambling dynamics, and it will spread out through the lattice. In previous studies of unit black holes, the focus was on operator growth in the fast-scrambling dynamics of a single cell. This paper is mostly about the complementary mechanisms of spatial growth on scales larger than $T^{-1}$. We can see the important phenomena by studying spatial lattices of low dimensional objects; namely qubits.

In this section, we will study precursors in a simple qubit system. In section \ref{first}, we will numerically simulate the system, and observe linear growth of the precursor as a function of time $t_w$. In section \ref{TN}, we will use this pattern of growth to qualitatively analyze the geometry of the minimal tensor network for a product of precursor operators.

\subsection{Precursor growth} \label{first}
The spin Hamiltonian we will use is an Ising system defined on a one-dimensional chain
\be
H = -\sum_i Z_i Z_{i+1} + g X_i + h Z_i, \label{ising-hamiltonian}
\ee
where $X_i,Y_i,Z_i$ are the Pauli operators on the $i$th site, $i = 1,2,...,n$. In our numerics, we use $n = 8$. We will consider two choices for the couplings: one for which the system is strongly chaotic ($g = -1.05$, $h = 0.5$) \cite{Hastings}, and one for which it is integrable ($g = 1$, $h = 0$).

We will study the size of the precursor associated to $Z_1$:
\be
Z_1(t_w) = e^{-iHt_w}Z_1e^{iHt_w}.
\ee
In this setting, where the operator begins at the endpoint of a one dimensional chain, size and radius are interchangeable. As a function of $t_w$, we define either as the number of sites $i$ such that $\tr [Z_1(t_w),A_i]^2$ is greater than or equal to one, with $A = X$, $Y$, or $Z$. The rate at which the operator grows can be controlled using the Lieb-Robinson bound for the commutator of local operators \cite{Lieb:1972wy,Hastings:2005pr,hastingslocality}. This bound states that
\be
\| [W_x(t_w),W_y] \| \le c_0\|W_x\| ~\|W_y\| ~ e^{c_1t_w - c_2|x-y|}, \label{lieb-robinson-bound}
\ee
where $c_0,c_1,c_2$ are constants that depend on the Hamiltonian. The norm is the operator (infinity) norm, and the bound is valid as long as the interactions decay exponentially (or faster) with distance. This bound implies that the radius of the operator can grow no faster than linearly, $r[Z_1(t_w)]<(c_1/c_2)t_w$.

It is not hard to see that some systems can saturate this linear behavior. A rather trivial example is the spin chain \eqref{ising-hamiltonian}, with $g=1$ and $h=0$. This system is integrable for all $g$ and can be solved by mapping to a system of free spinless Majorana fermions via the nonlocal Jordan-Wigner transformation (see e.g. \cite{sachdev2011quantum}). This nonlocal mapping relates $a_k$, the fermion annihilation operator at site $k$, to spin operators of the form $X_1X_2...X_{k-1}Z_k$ and $X_1X_2...X_{k-1}Y_k$. The fact that the free fermions propagate linearly in time corresponds, in the spin variables, to a linear growth of the operator.\footnote{This is an interesting case where growth and movement are related, despite footnote \ref{growthvsmovement}. The reason this is possible is that the change of variables is nonlocal.}

\begin{figure}[ht]
\begin{center}
\includegraphics[scale=.35]{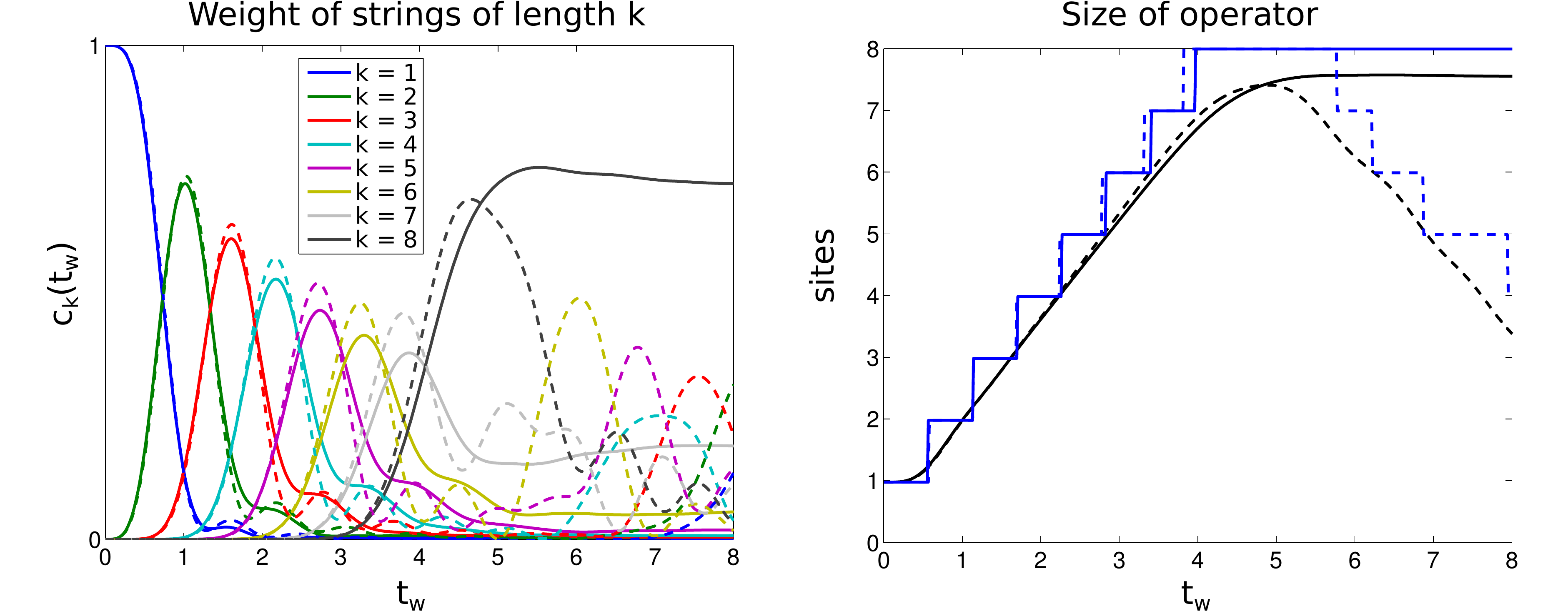}
\end{center}
\caption{Ballistic growth of the operator $Z_1(t_w)$, evolved with the chaotic $g =-1.05$, $h = 0.5$ Hamiltonian (solid) and the integrable $g = 1$, $h = 0$ Hamiltonian (dotted). {\bf Left:} $c_k(t)$ is the sum of the squares of the coefficients of Pauli strings of length $k$ in $Z_1(t_w)$. Notice that the integrable and chaotic behavior is rather similar until the strings grow to reach the end of the chain ($n=8$ spins). {\bf Right:} for both types of evolution, the size grows linearly until it approaches the size of the system. After this point, the chaotically-evolving operator saturates, while the integrably-evolving operator begins to shrink. The blue ``staircase'' curves show the size $s[Z_1(t_w)] $. The smooth black curves show $s_2[Z_1(t_w)]  \propto \sum_k k \ c_k(t_w)$.}
\label{fig-spin-chain-growth}
\end{figure}

This integrable system is clearly very special: even typically diffusive quantities such as energy density move ballistically. One might naively guess that the linear precursor growth is also exceptional, and that a chaotic system will exhibit slower (perhaps diffusive) growth. To show that this is not the case, we numerically analyze a chaotic version ($g = -1.05$, $h = 0.5$) of the same spin chain alongside the integrable model, and plot the size of the precursor in the right panel of Fig.~\ref{fig-spin-chain-growth}. The size, according to the commutator definition, corresponds to the staircase plots. The rate of growth is clearly linear for both the chaotic (solid blue) and integrable (dashed blue) curves. The only significant difference occurs once the operator grows to the size of the entire chain---in the chaotic system it saturates and in the integrable system it begins to shrink.

The commutator is a useful measure of the size of the operator, but having an explicit numerical representation allows us to understand the growth in other ways. It is helpful to think about expanding $Z_1(t_w)$ in a basis of Pauli strings, e.g. $X_1Z_2Y_3I_4X_5X_6Y_7Z_8Y_9\dotso$. Starting from the simple operator $Z_1$, such strings are generated by the Baker-Campbell-Hausdorff formula
\be
Z_1(t_w) = Z_1 - it_w~[H,Z_1] - \frac{t_w^2}{2!}~[H,[H,Z_1]] + \frac{it_w^3}{3!}~[H,[H,[H,Z_1]]] + \dots ~. \label{BCH-formula}
\ee
For example, suppressing coefficients and site indices, one finds the sixth order term
\begin{align}
 [H, [H, [H,[H, [H, [H, Z]]]]] =&X, ~~Z, ~~XX, ~~XZ, ~~YY, ~~ZX, ~~ZZ, ~~IX,~~XXX, ~~XXZ, \notag \\
 &XYY, ~~XZZ, ~~YXY, ~~YYZ, ~~ZXZ, ~~XXXZ. \label{chaotic-commutator}
\end{align}
As $t_w$ increases, high order terms in the BCH expansion become important, corresponding to longer and more complicated Pauli strings.

To quantify this growth, we will group together Pauli strings according to their length. We define the length of a Pauli string as the highest site index that is associated to a non-identity Pauli operator. For example, the length of $X_1I_2Y_3$ is three. Given a Pauli string representation of an operator, we define $c_k$ as the sum of the squares of coefficients of all Pauli strings of length $k$. We plot the $\{c_k\}$, as a function of $t_w$, in the left panel of Fig.~\ref{fig-spin-chain-growth}. Again, the chaotic and integrable systems track each other well until the operator becomes roughly as large as the entire system. Using these coefficients, we can define a smoother notion of size by taking
\be
s_2[Z_1(t_w)] = \frac{\sum_k k \ c_k(t_w)}{\sum_k c_k}.
\ee
This is plotted as the black curves in the right panel of Fig.~\ref{fig-spin-chain-growth}. It agrees fairly well with the staircase definition using the commutator. The initial delay in the growth is due to the fact that $Z_1$ needs to be converted to $Y_1$ before the strings can grow in length. This can be thought of as the time to ``scramble'' a single site.

\subsection{TN for multiple localized precursors}\label{TN}
\begin{figure}[t]
\begin{center}
\includegraphics[scale=.7]{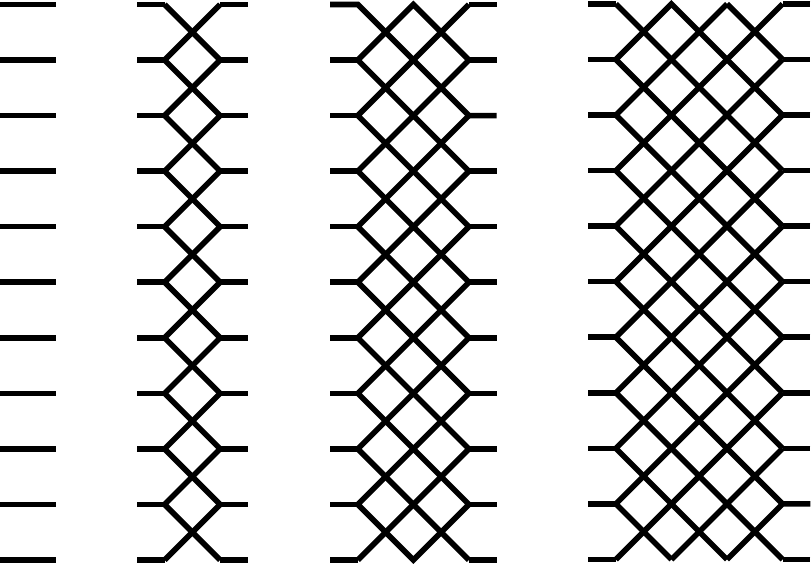}
\end{center}
\caption{The tensor network description of the identity operator (left) and operators $e^{-iHt}$ with successively larger $t$. }
\label{fig-tensor-network1}
\end{figure}

In this section, we will assume that precursor operators grow at a rate $v_B$. That is, $r[W_x(t_w)] = v_B t_w$. We will use this pattern of growth to characterize the tensor networks associated to a product of localized precursors. To begin, let us review the TN geometry associated to the time evolution operator $e^{-iHt}$. This is illustrated in Fig.~\ref{fig-tensor-network1} for a quantum system of eleven sites.
 Each diagram represents a formula for $[e^{-iHt}]_{mn}$, with different values of $t$. The line endpoints on the left represent the tensor decomposition of the $m$ index into eleven indices $m_1,...,m_{11}$, and the line endpoints on the right represent the $n$ index as $n_1,...,n_{11}$. Line segments correspond to Kronecker delta contractions. Thus the figure at the far left is a formula for the identity matrix: $I_{mn} = \delta_{m_1n_1}\delta_{m_2n_2}...\delta_{m_{11}n_{11}}$.

Intersections of lines represent a tensor with rank equal to the number of lines. In the figure, we have only four-fold intersections, corresponding to tensors $t_{i_1i_2i_3i_4}$. The grids of intersecting lines represent a particular contraction of a large number of these tensors. It is known that time evolution by a local Hamiltonian can be represented in terms of such networks, using a technique known as Trotterization.

We have presented the TN as giving a formula for an operator, but we can also think about it as a formula for an entangled pure state of two quantum systems $L$ and $R$. In this interpretation, the left ends correspond to indices in the $L$ system, and the right ends correspond to indices in the $R$ system. Contracting the tensors gives the wave function. This is an example of a general correspondence: an operator $A_{mn}$ can be understood as the wave function for a state given by acting with the operator $A$ on one side of a maximally entangled state: $|A\rangle = A\sum_{i}|i\rangle|i\rangle = \sum_{mn}A_{mn}|m\rangle|n\rangle$.

Thought about either way, the basic feature of this network is that it grows linearly with time. The work of \cite{Hartman:2013qma} pointed out the relationship between this linear growth and the geometry of the Einstein-Rosen bridge of the time-evolved thermofield double state $e^{-iHt}|TFD\rangle$.

\subsubsection*{One precursor}
\begin{figure}[t]
\begin{center}
\includegraphics[scale=.8]{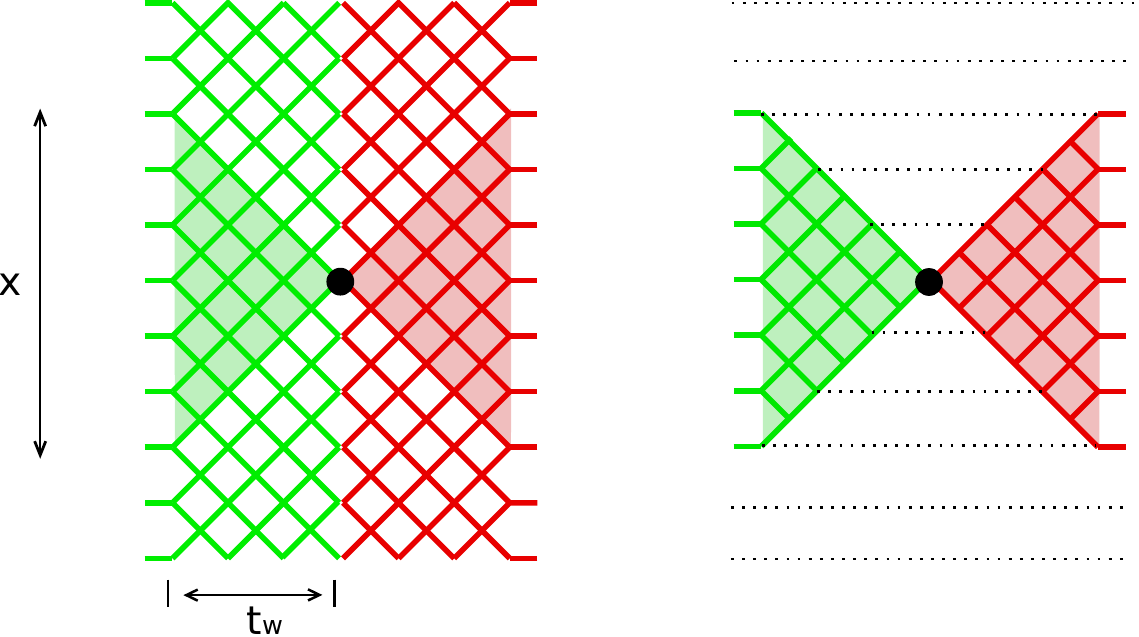}
\end{center}
\caption{{\bf Left:} the naive tensor network describing a precursor operator $e^{-iHt_w}We^{iHt_w}$. The red network represents backwards time evolution, the black dot represents the local operator $W$, and the green network represents forwards time evolution. Shading indicates the region affected by the linearly growing $W$ insertion. In the unshaded region, the forwards and backwards evolutions cancel. {\bf Right:} the network after removing tensors that cancel. The dotted lines indicate contractions; their endpoints should be identified.}
\label{fig-tensor-network}
\end{figure}
Let us now understand the tensor network for a single precursor operator $W_x(t_w)$. The structure is illustrated in Fig.~\ref{fig-tensor-network}. Concatenation of tensor networks represents multiplication of operators, and in the left panel, we have a naive TN for the operator, in which we simply concatenate the networks for $e^{-iHt_w}$, $W_x$, and $e^{iHt_w}$. The tensor network for the $W$ operator is the identity on all sites except for the central one, represented by the dot. This naive tensor network is an explicit representation of a time-fold \cite{Heemskerk:2012mn,Susskind:2013lpa}, where we evolve backwards in time, insert the operator, then evolve forwards again. As in \cite{Stanford:2014jda}, we can simplify the network by considering the partial cancellation between $e^{-iHt_w}$ and $e^{iHt_w}$. If we had not inserted the $W_x$ operator, the cancellation would be complete. This means that we can remove the tensors in the region that is not affected by the linearly growing precursor. After doing so, we obtain the simpler network shown in the right panel of Fig.~\ref{fig-tensor-network}.\footnote{D.S. is grateful to Don Marolf for pointing out an error in a previous version of this argument.}

\begin{figure}[t]
\begin{center}
\includegraphics[scale=.7]{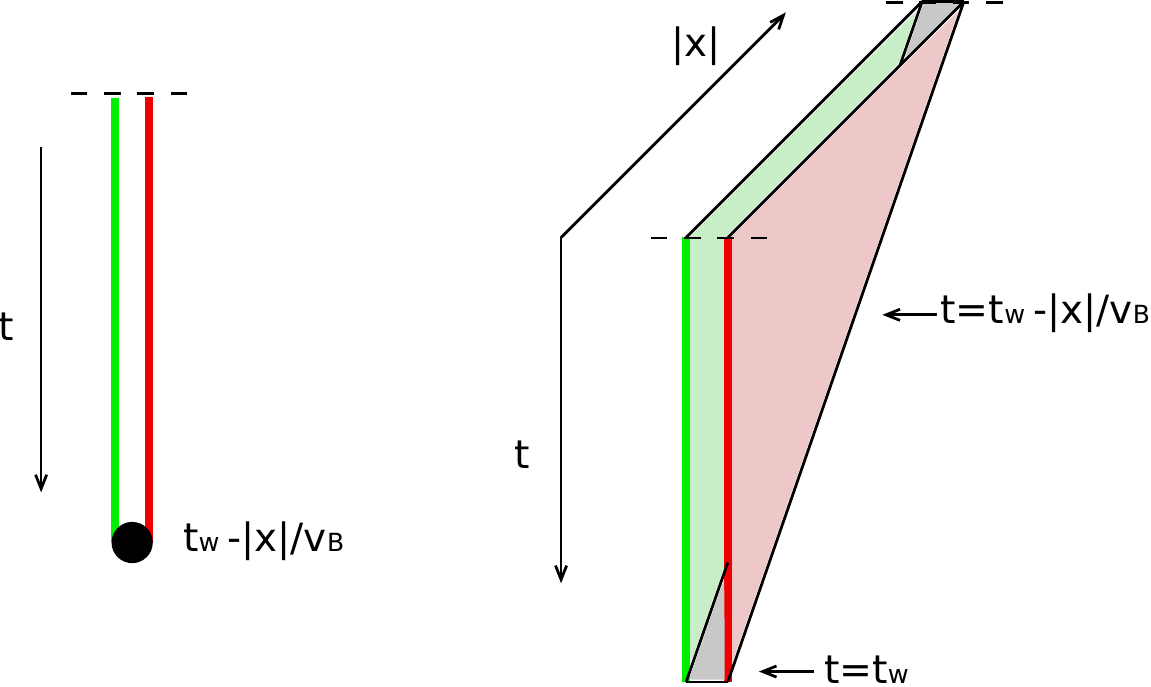}
\end{center}
\caption{Fibering the position-dependent time fold over the $x$ space gives the minimalized TN. The geometry at right is equivalent to the right panel of Fig.~\ref{fig-tensor-network}.}
\label{fig-time-fold}
\end{figure}

This ``minimalized'' tensor network can be understood in terms of a position-dependent time-fold. Because the region of influence of the operator grows outwards at a rate $v_B$, at distance $|x|$ from the insertion of the operator, we only need to include a time fold of length $t_w - |x|/v_B$; the rest of the fold cancels. The minimalized TN geometry in Fig.~\ref{fig-tensor-network} can be constructed as the fibration of this position-dependent fold over the $x$ space, as shown in Fig.~\ref{fig-time-fold}. This procedure extends to higher dimensions, where one finds a geometry consisting of two solid cones, glued together along their slanted faces.

\subsubsection*{Multiple precursors}

Now, consider a product of $n$ localized precursor operators\footnote{We emphasize that the times $t_i$ are not necessarily in time order. In fact, the most interesting case is the one in which the differences between adjacent times alternate in sign.}
\be
 W_{x_{n}}(t_{n}) \dots W_{x_1}(t_1) = e^{-iHt_n} W_{x_n} e^{iHt_n}  \dots e^{-iHt_1} W_{x_1} e^{iHt_1}.
\ee
 As before, we can form a naive TN by simply concatenating the networks for each of the $W$ and $e^{\pm iHt}$ operators on the RHS. To form the minimalized TN, we cancel adjacent regions of forward and backward evolution outside the influence of any insertion. This procedure is explicit, but it becomes complicated as the number of operators increases, as their insertion positions are varied, and for systems with spatial dimension larger than one. Even with three operators, the geometry can be rather nontrivial, as shown in\begin{figure}[t]
\begin{center}
\includegraphics[scale=.75]{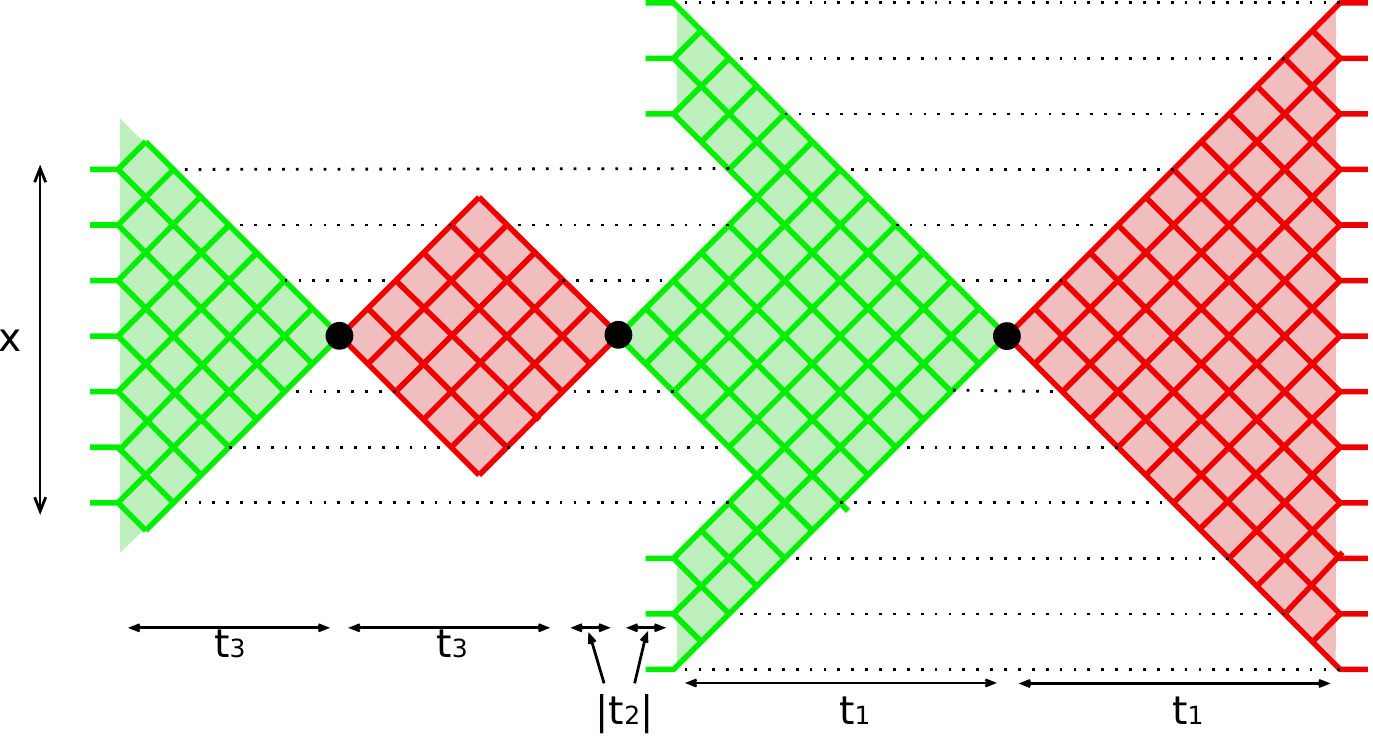}
\end{center}
\caption{The minimal tensor network describing a product of three precursor operators $\mathcal{O} = W(t_3) W(t_2) W(t_1)$ inserted at the same spatial point and with $0<-t_2<t_3<t_1$.}
\label{fig-three-precursors}
\end{figure} Fig.~\ref{fig-three-precursors}.

In order to compare with holography, it will be useful to emphasize the representation of the minimalized TN using position-dependent time-folds. 
\begin{figure}[t]
\begin{center}
\includegraphics[scale=.9]{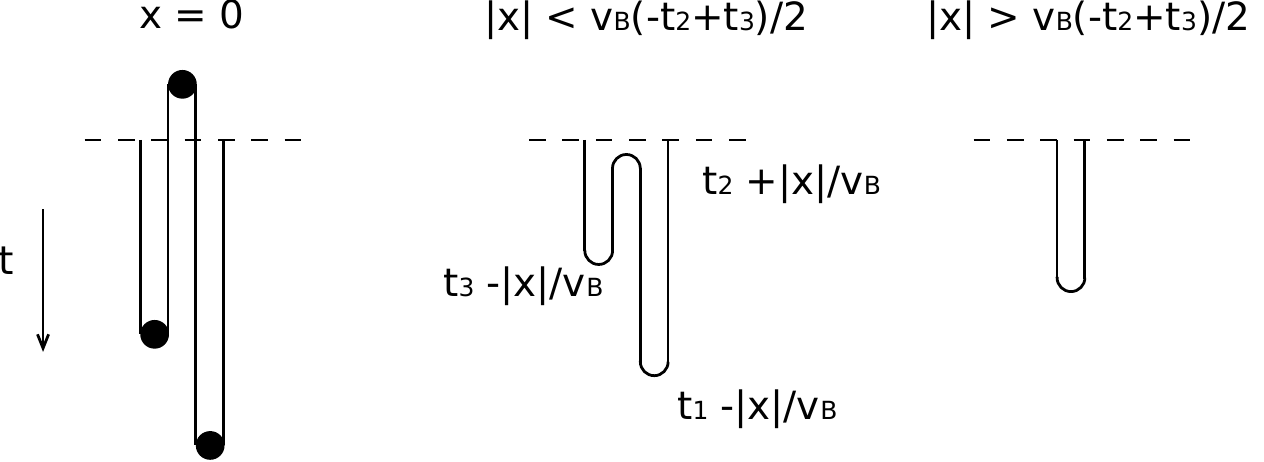}
\end{center}
\caption{Slices of the $x$-dependent time-fold for the product of three local precursor operators inserted at $x=0$ with $0<-t_2<t_3<t_1$. As $|x|$ increases, the folds are pulled inwards. Eventually the second and third folds merge, leaving a single fold.}
\label{fig-time-folds}
\end{figure} Let us begin with the case where all operators are localized at $x = 0$. We define the position-dependent time fold as follows. At $x = 0$, we simply have the folded time axis defined by the times $t_1,...,t_n$, as in the spherically symmetric case studied in \cite{Stanford:2014jda}. Depending on the configuration of times, this will have $k \le n$ folds, and $(n-k)$ through-going insertions (see \cite{Stanford:2014jda}). As $|x|$ increases, each fold gets pulled inwards, by an amount $|x|/v_B$; that is, the time location of the fold associated to operator $j$ becomes $t_j \pm |x|/v_B$, with a sign that depends on the direction of the fold. This reflects the fact that as we move farther from the insertion, there is additional cancellation between forwards and backwards time evolution. At certain special values of $|x|$, pairs of folds will merge and annihilate, leaving behind two fewer folds (Fig.~\ref{fig-time-folds}). This procedure defines a folded time axis as a function of $x$. Fibering this geometry over the $x$ space gives the geometry of the minimalized tensor network.

In the case where the operators are localized at general positions $\{x_j\}$, the procedure is slightly more complicated. To define the position-dependent fold at location $x$, we begin with the position-independent time fold, passing through insertions at times $\{t_j\}$. We then replace each $t_j$ by a new variable $\tau_j$, and we minimize the length of the folded time axis subject to two constraints: first, the time fold must pass through each $\tau_i$ in order; second, the $\{\tau_i\}$ must satisfy $|\tau_j - t_j| \le |x - x_j|/v_B$. The minimalized TN is the fibration of this time fold over the $x$ space. In general, it consists of flat regions glued together at the loci of the folds.

\section{Holographic systems}\label{2}
In this section, we will use holography to study the dynamics of precursors. The analysis will be highly geometric: the action of a precursor on the thermofield double state generates a gravitational shock wave that distorts the Einstein-Rosen bridge, as in \cite{Shenker:2013pqa}. By analyzing the transverse profile of this shock wave, in section \ref{2.2}, we will be able to estimate the commutator with other operators, and thus the size of the precursor as a function of time. In section \ref{2.3}, we will study the geometry dual to a product of local precursor operators. We will find a detailed match with the geometry of the corresponding tensor networks, characterized in section \ref{TN}.

\subsection{Localized shock waves} \label{2.1}
Let us start by reviewing the AdS black hole dual to the thermofield double state $|TFD\rangle$ of two CFTs $L$ and $R$ \cite{Maldacena:2001kr}.\footnote{We should clarify the role of the thermofield double state and the background black hole geometry. Why couldn't we study the same problem with vacuum AdS as the background? The answer is that matrix elements of the precursor, and its commutators with other operators, depend on the energy. Near the vacuum, the dynamics is integrable. The relevant commutators might be large as operators, but they have small expectation value in the vacuum state (this can be seen explicitly in the spin model numerics). The black hole geometry allows us to study nontrivial matrix elements.} Planar AdS black holes have only one scale, namely $\ell_{AdS}$, and we can write the metric in terms of dimensionless coordinates as
\begin{align}
ds^2 &= \ell_{AdS}^2\Big[-f(r)dt^2 + f^{-1}(r)dr^2 + r^2 dx^idx^i\Big] \\
&\hspace{50pt}f(r) = r^2 - r^{2-d},
\end{align}
where $i$ runs over $(d-1)$ transverse directions. It will convenient to use the smooth Kruskal coordinates, $u$ and $v$, which are defined in terms of $r$ and $t$ by 
\begin{align}
uv=-e^{f'(1)r_*(r)}, \qquad   u/v = -e^{-f'(1)t}, \label{kruskal-def}
\end{align}
where the tortoise coordinate is $r_*(r)=\int_\infty^r dr' f^{-1}(r')$. In terms of these Kruskal coordinates, the black hole metric can be written
\begin{align}
ds^2 = \ell_{AdS}^2\Big[-A(uv) dudv + B(uv) dx^idx^i\Big]\\
A(uv) = -\frac{4}{uv}\frac{f(r)}{f'(1)^2} \hspace{20pt} B(uv) = r^2.
\end{align}
The horizon of the black hole is at $r = 1$, or $uv = 0$. The inverse temperature is $\beta =  4\pi/d$.

Following \cite{Shenker:2013pqa}, we will act on the thermofield double state with a precursor of an operator in the $L$ CFT:
\be
W_x(t_w)|TFD\rangle = e^{-iH_Lt_w}W_x \, e^{iH_Lt_w}|TFD\rangle,
\ee
where the $W_x$ operator is an approximately local, thermal scale operator acting near location $x$ on the left boundary. We would like to understand the geometry dual to this state.\footnote{In order to have a well defined notion of geometry, we will should consider a coherent operator $W$ built out of a large but fixed number (e.g. 100) of quanta. However an operator corresponding to even a single quantum will lead to similar effects on the commutator that we estimate below; the metric should be understood in the single-particle case as giving the eikonal phase \cite{Amati:1987uf}.} If the time $t_w$ is not large, then the geometry will not be substantially affected by the perturbation. However, Schwarzschild time evolution acts near the horizon as a boost, and as we make the Killing time $t_w$ of the peturbation earlier, its energy in the $t = 0$ frame gets boosted $\propto e^{2\pi t_w/\beta}$. When this exceeds the $G_N$ suppression from the gravitational coupling, $t_w \sim t_* = \frac{\beta}{2\pi}\log N^2$, the geometry will be affected.\footnote{This potential important of this time scale was first noticed by \cite{Schoutens:1993hu}. The connection to scrambling was made in \cite{Hayden:2007cs}, and the connection to gauge/gravity duality was made in \cite{Sekino:2008he}. This connection was made precise in \cite{Shenker:2013pqa}.}

If $G_N$ is small, then the boost must be large in order to overcome the suppression. The associated stress energy distribution is highly compressed in the $u$ direction and stretched in the $v$ direction. We can replace this by a stress tensor localized on the $u = 0$ horizon,
\be
T_{uu} = \frac{E}{\ell_{AdS}^{d+1}} e^{2\pi t_w/\beta}\delta(u) a_0(x),
\ee
where $E$ is the dimensionless asymptotic energy of the perturbation, and $a_0$ is a function concentrated within $|x|\lesssim 1$, with integral of order one. The precise form of $a_0$ depends on details of the perturbation, as well as the propagation to the horizon. 

\begin{figure}[ht]
\begin{center}
\includegraphics[scale=.9]{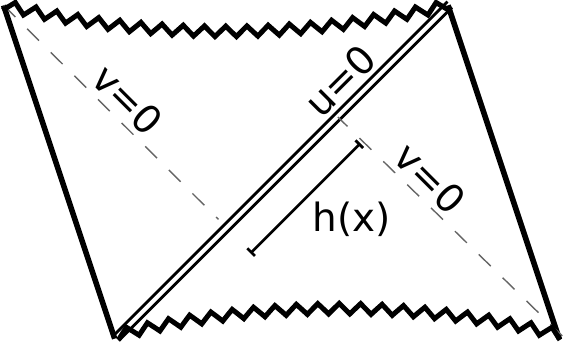}
\end{center}
\caption{A constant $x$ slice through the one-shock geometry. The geometry consists of two halves of the eternal AdS black hole, glued together with a null shift of magnitude $h(x)$ in the $v$ direction. The shock lies along the surface $u = 0$. In the coordinates (\ref{shockAnsatz}), the surface $v = 0$ is discontinuous.}
\label{oneshock}
\end{figure}
The backreaction of this matter distribution is extremely simple. It was worked out first in \cite{Dray:1984ha} for the case of a flat space Schwarzschild black hole, and by \cite{Sfetsos:1994xa,Shenker:2013pqa} for AdS black holes. The idea is to consider a shock wave ansatz of the form\footnote{For finite $t_w$, the metric will not be exactly of this type. Corrections are analyzed in appendix \ref{appendixC}.}
\be
ds^2 = \ell_{AdS}^2\Big[-A(uv) du dv + B(uv) dx^idx^i + A(uv)\delta(u)h(x)du^2\Big].\label{shockAnsatz}
\ee
This metric can be understood as two halves of the AdS black hole, glued together along $u = 0$ with a shift of magnitude $h(x)$ in the $v$ direction (see Fig.~\ref{oneshock}). By evaluating the curvature of this metric (setting $u \delta'(u) = -\delta(u)$ and $u^2\delta(u)^2 = 0$) and plugging into Einstein's equations, we find a solution if
\be
\left(-\partial_i\partial_i + \mu^2\right)h(x) = \frac{16\pi G_N}{A(0)\ell_{AdS}^{d-1}}E e^{\frac{2\pi}{\beta}t_w}a_0(x), \label{equation-for-h}
\ee
where $\mu^2 = \frac{d(d-1)}{2}$. For $|x|\gg 1$, the solution to this equation will depend only on the integral of $a_0(x)$, which we can replace with a delta function. The differential operator in Eq.~(\ref{equation-for-h}) can then be inverted in terms of Bessel functions. Expanding for large $|x|$, and assuming a thermal-scale initial energy $E$, we find the solution
\be
h(x) = \frac{e^{\frac{2\pi}{\beta}(t_w - t_*)-\mu |x|}}{{|x|^{\frac{d-2}{2}}}} \ , \label{shock-solution}
\ee
where the scrambling time $t_* = \frac{\beta}{2\pi}\log \frac{c\ell_{AdS}^{d-1}}{G_N} \approx \frac{\beta}{2\pi} \log N^2$ has been defined with $c$ chosen to absorb certain order-one constants. For $|x| \lesssim 1$, the approximation of $a_0$ by a delta function is incorrect; the power-law singularity in the denominator should be smoothed out.

The strength of the shock wave is exponentially growing as a function of $t_w$, reflecting the growing boost of the initial perturbation. However, it is exponentially suppressed as a function of $x$. In the next section, we will use the interplay of these exponentials to determine the growth of the precursor operator $W_x(t_w)$.

\subsection{Precursor growth}\label{2.2}
To measure the size of the precusor, let us consider the expectation value of the commutator-squared
\be
C(t_w,|x-y|) = \tr\left\{\rho(\beta)[W_x(t_w),W_y]^\dagger [W_x(t_w),W_y]\right\},
\ee
where $\rho(\beta)$ is the thermal density matrix. To simplify the analysis, we will consider the setting in which {\it (i)} $W_x$ and $W_y$ are both unitary operators, so that $0\le C \le 2$, and {\it (ii)} they correspond to different bulk fields. As before, we will define the radius of the operator at time $t_w$ as the maximum distance $|x-y|$ such that $C(t_w,|x-y|)$ is equal to one.

To calculate the above expectation value we will follow the procedure used by Ref.~\cite{Shenker:2013yza} in the spherical shock case. Similar calculations were also previously done by \cite{'tHooft:1990fr,Kiem:1995iy}. Using the purification by the thermofield state, we can write
\begin{align}
C(t_w,|x-y|) &= \langle TFD|[W_x(t_w),W_y]^\dagger[W_x(t_w),W_y]|TFD\rangle \\
&=2 - 2 Re\langle \psi|\psi'\rangle,
\end{align}
where $|\psi\rangle = W_x(t_w)W_y|TFD\rangle$ and $|\psi'\rangle = W_yW_x(t_w)|TFD\rangle$. It is helpful to think about this inner product in the $t = 0$ frame, in which the $W_x$ perturbation is highly boosted, and the $W_y$ perturbation is not. Let us suppose that $t_w$ is not too large, so the relative boost is a small fraction of the Planck scale. Then the $W_x$ operator creates a mild shock wave, and the $W_y$ operator creates a field theory disturbance that propagates on this background. The difference between applying the $W_y$ operator before or after the $W_x$ operator is a null shift in the $v$ direction of magnitude $h(y-x)$ \cite{Shenker:2013yza}. The inner product of the states $|\psi\rangle$, $|\psi'\rangle$ then reduces to the following question: if we take the field theory state corresponding to the $W_y$ perturbation and shift it by $h(x-y)$ in the $v$ direction, what is its overlap with the original unshifted state?

If the strength of the shock is sufficiently weak, the shift is small and the overlap is close to one; we recover a small commutator. However, once $h(x-y)\sim 1$ the shift becomes of order the typical wavelength in the field theory excitation created by $W_y$, and the inner product begins to decrease. Because the strength of the shock is exponential in $t_w$, the overlap will be quite small within a few thermal times. At order one precision, we can therefore determine the size of the operator by setting the formula (\ref{shock-solution}) equal to one. We find (for perturbations with order one energy $E$)
\begin{align}
r[W_x(t_w)] &= \frac{2\pi}{\beta\mu}t_w - \frac{1}{\mu}\log N^2 - O(\log t_w) \\
&= v_B (t_w - t_*) - O(\log t_w),
\end{align}
where $t_* = \frac{\beta}{2\pi}\log N^2$ and \cite{Shenker:2013pqa}
\be
v_B = \frac{2\pi}{\beta\mu} = \sqrt{\frac{d}{2(d-1)}}.
\ee
Here, $d$ is the space time dimension of the boundary theory. The radius is negative for $t_w$ less than the scrambling time $t_*$, indicating that the commutator is small everywhere. However, for larger values of $t_w$, the region of influence spreads ballistically, with the velocity $v_B$. This is the speed at which precursors grow, or equivalently, the speed at which the butterfly effect propagates.

It is interesting to compare this speed with $v_E$, the rate at which entanglement spreads. On rather general grounds, entanglement should spread no faster than the commutator of local operators.\footnote{We are grateful to Sean Hartnoll for emphasizing this point to us.} The speed $v_E$ was recently computed by \cite{Hartman:2013qma,Liu:2013iza} for holographic systems dual to Einstein gravity, with the result
\be
v_E = \frac{\sqrt{d}(d-2)^{\frac{1}{2} - \frac{1}{d}}}{[2(d-1)]^{1 - \frac{1}{d}}}.
\ee One can check that $v_B \ge v_E$, with equality at $d = 1+1$.

The speed $v_B$ will be corrected in bulk theories that differ from Einstein gravity. Stringy effects are considered in \cite{Shenker:2014cwa}. In appendix \ref{GB}, we work out the shock solutions in Gauss-Bonnet gravity\footnote{We are grateful to Steve Shenker for suggesting this.}. The result for $d \ge 4$ is that the velocity is corrected to
\begin{align}
v_B(\lambda_{GB})&=\frac{1}{2}\sqrt{1+\sqrt{1-4\lambda_{GB}}}\sqrt{\frac{d}{d-1}}\\
&=\left(1 - \frac{\lambda_{GB}}{2} +...\right)v_B.
\end{align}
This speed increases for negative $\lambda_{GB}$, and it exceeds the speed of light (in $d = 4$) for $\lambda_{GB} < -3/4$. In fact, Gauss-Bonnet gravity is known to violate boundary causality for $\lambda_{GB} < -0.36$ \cite{Hofman:2008ar}. (In fact the recent work of \cite{Camanho:2014apa} shows that maintaining causality requires an infinite number of massive higher spin fields for any nonzero value of $\lambda_{GB}$.)

\subsection{ERB dual to multiple localized precursors}\label{2.3}
In this section, we will characterize the geometry dual to a product of localized precursor operators,
\be
W_{x_n}(t_n)...W_{x_1}(t_1)|TFD\rangle.\label{what'sup}
\ee 
Following \cite{Dray:1985yt}, Ref.~\cite{Shenker:2013yza} showed how to construct states of this type in the spatially homogeneous case, building the geometry up one shock at a time. If the masses of the perturbations are small and the times are large, the geometry consists of a number of patches of AdS-Schwarzschild, glued together along their horizons, with null shifts $h_1...h_n$ determined by the times $t_1...t_n$. For spatially localized perturbations, there are two differences: first, the null shifts depend on the transverse position $h_{1}(x)...h_n(x)$ and second, the geometry after shocks collide is not generally known. These regions will not substantially affect our analysis, for reasons explained below.

The point we will emphasize is that the intrinsic geometry of the maximal spatial slice\footnote{The timelike interval inside the ERB remains of order $\ell_{AdS}$, even as the shocks make the total spatial volume very large. The conjecture of \cite{maldacenaTN} was that the TN geometry reflects a coarse-graining of the ERB on scale $\ell_{AdS}$. At this level, the maximal spatial surface represents the entire ERB.} through the ERB, $\Sigma_{max}$, agrees with the geometry of the corresponding TN on scales large compared to $\ell_{AdS}$. In general, the geometries dual to (\ref{what'sup}) do not have any symmetry, and finding the maximal surface exactly would require the solution of a nonlinear PDE. In order to understand the large-scale features of $\Sigma_{max}$, we use the following fact: in the exact AdS-Schwarzschild geometry, maximal surfaces within the ERB are attracted to a spatial slice defined by a constant Schwarzschild radius $r = r_m$. This was first pointed out (for co-dimension two surfaces) in \cite{Hartman:2013qma}, and we will refer to the attractor surface as the Hartman-Maldacena (HM) surface. 

Within each patch of a multi-shock wormhole, the maximal surface hugs the HM surface of that patch. As $\Sigma_{max}$ passes through a shock connecting adjacent patches, it transitions from one HM surface to the next. Due to the lack of symmetry, we will not be able to determine $\Sigma_{max}$ in this transition region. However, the transition takes place over an intrinsic distance of order $\ell_{AdS}$. Each HM surface is intrinsically flat, so the surface $\Sigma_{max}$ consists of approximately flat regions, joined by curved regions of size $\sim\ell_{AdS}$. The flat regions grow large in proportion to the time between $W$ insertions. We will calculate their size and shape, in the limit that they become large, and match to the geometry of the TN associated to (\ref{what'sup}).

Before we begin, let us make one more technical comment. In the analysis below, we will focus on the geometry of the ``decoupled maximal surface'' $\Sigma_{dec}$, which maximizes a volume-like functional, $V_{dec}$, obtained from the true volume by dropping terms involving $x$ gradients. Finding this surface is technically simpler, but we argue in appendix \ref{B} that $\Sigma_{dec}$ and $\Sigma_{max}$ agree at the level of an $\ell_{AdS}$ coarse-graining, becuase both are attracted to the same HM surfaces.

\subsubsection*{A simple example}
\begin{figure}[t]
\begin{center}
\includegraphics[scale=.8]{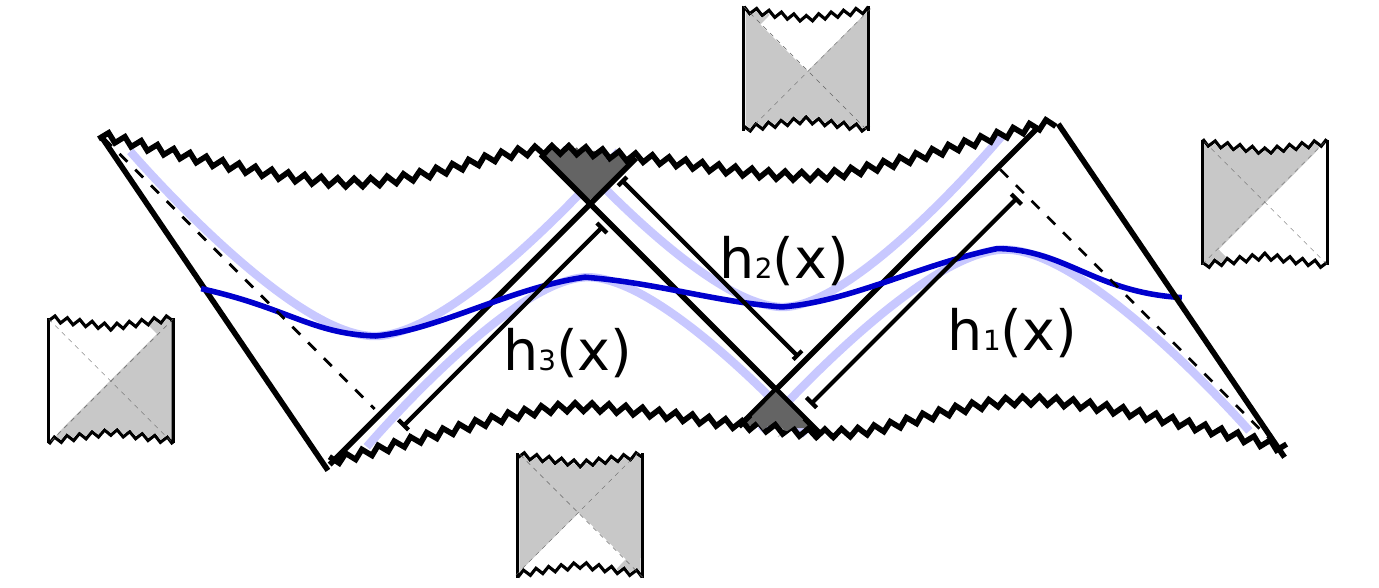}
\end{center}
\caption{A slice through a three-shock geometry. The white pre-collision regions are cut and pasted from the displayed white regions of AdS black holes. The pale blue curves are HM surfaces of the separate patches. The dark blue curve is the maximal surface. Dark grey indicates post-collision regions. As the shocks grow stronger, the dark grey regions shrink and the maximal surface tracks more of each HM surface.}
\label{kinked}
\end{figure}

We will start by demonstrating agreement with the TN geometry in a case in which {\it (i)} all shocks are centered near $x_j = 0$, {\it (ii)} all odd numbered times $t_1,t_3,...$ are positive and all even nubered times $t_2,t_4,...$ are negative, and {\it (iii)} all shocks are strong, i.e. adjacent time differences are large, $|t_{j+1}-t_j|  - 2t_* \gg 1$. The $x = 0$ slice through the geometry dual to a configuration of this type is shown in Fig.~\ref{kinked}. Across shock $j$ (counting from the right), we have a null shift of magnitude
\be
h_j(x) = \frac{e^{\frac{2\pi}{\beta}(\pm t_j - t_*)-\mu |x|}}{{|x|^{\frac{d-2}{2}}}}.\label{profile}
\ee
The upper sign is appropriate for the $v$ shifts associated to odd-numbered perturbations (right-moving shocks), and the lower sign is appropriate for $u$ shifts associated to the even-numbered perturbations (left-moving shocks). This form is accurate for $|x|\gg 1$. At smaller values of $|x|$, the singularity in the denominator should be smeared out over the thermal scale $\sim \ell_{AdS}$.

Because the defining functional for $\Sigma_{dec}$ does not contain $x$ gradients, the surface can be constructed independently at each $x$, by solving a maximal surface problem in a spatially homogeneous shock geometry with $x$-independent shifts $h_{i} = h_{i}(x)$. This problem was studied in \cite{Stanford:2014jda}, following \cite{Hartman:2013qma}. For a configuration with $n$ shocks, the intersection of $\Sigma_{dec}$ with $x = 0$ is a curve made up of $n+1$ segments. Two of these connect to the asymptotic boundaries, and $n-1$ of them pass between shocks. All but an order-one contribution to the length of these segments comes from regions near the HM surface. Following the analysis in \cite{Stanford:2014jda} one finds that the large $t$ behavior of the length of the $j$-th segment is
\be
\log h_{j+1}h_j  \propto  |t_{j+1}-t_j| - 2t_*.
\ee
These segments can be identified with pieces of the folded time axis at $x = 0$.

So far, this is identical to the homogeneous case in \cite{Stanford:2014jda}. But now, keeping the same configuration of shocks, we consider a slice at nonzero $x$. As $|x|$ increases, the only difference will be that the shocks are weaker, according to the transverse profile in Eq.~(\ref{profile}), and the segments will be correspondingly shorter,
\be
|t_{j+1}-t_j| - 2t_* - 2|x|/v_B - O(\log |x|).
\ee Apart from the $t_*$ and the logarithm, this agrees with the $|x|$ dependence of the position-dependent folds from section \ref{TN}. The $t_*$ represents further cancellation of $e^{-iHt}$ and $e^{iHt}$ during single-site fast scrambling \cite{Stanford:2014jda}. The logarithm in $|x|$ indicates a slight modification of linear growth, but it is subleading in the limit that the segments are large. An important point is that, even as $x$ is varied, all but an order one contribution to the length will come from the region near the flat HM surfaces, thus all but an order-one pieces of $\Sigma$ will be approximately flat. Varying $x$, the geometry of $\Sigma_{dec}$ is a fibration of the collection of segments, that is, a fibration of the folded time axis.

\begin{figure}
\begin{center}
\includegraphics[scale=.6]{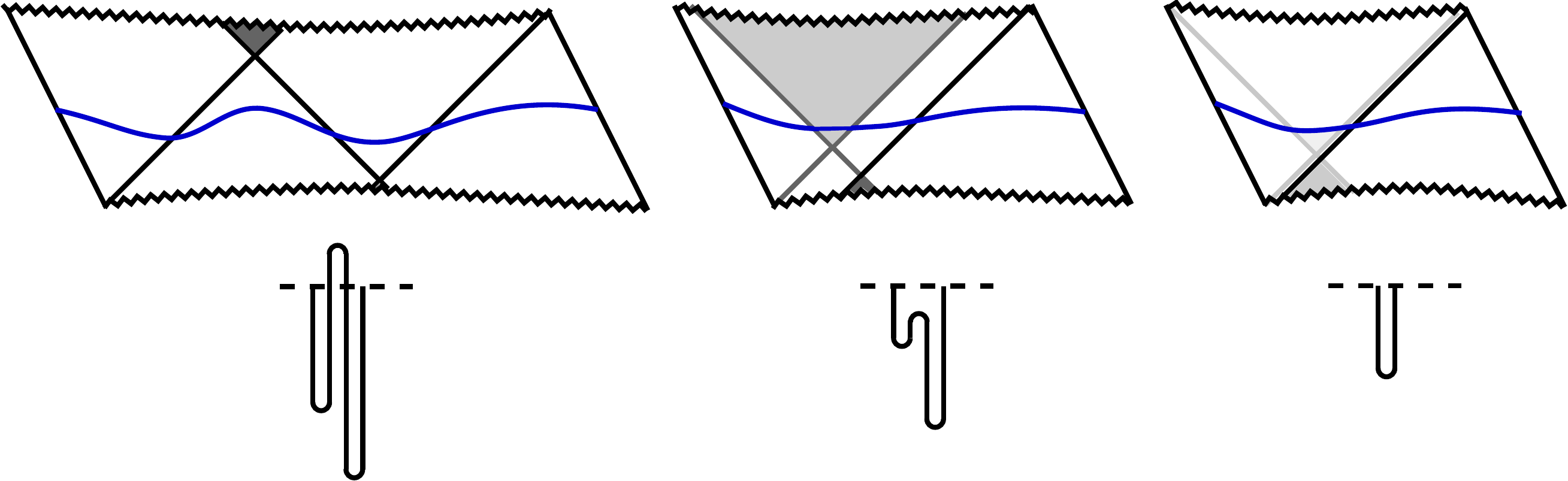}
\end{center}
\caption{The ERB implementation of the transition in Fig.~\ref{fig-time-folds}. At left, both $h_1 h_2$ and $h_2h_3$ are large, so the shaded post-collision regions are far from $\Sigma$. Increasing $|x|$, the shocks become weaker. Eventually $h_2h_3$ becomes order one (middle) and the surface passes through a nonlinear post-collision region. At yet larger $|x|$, $h_2 h_3 \ll 1$; nonlinear effects are small and the geometry has effectively one shock (right). The transition occurs over an order $\ell_{AdS}$ region in $\Sigma$.}
\label{transition}
\end{figure}

This analysis is enough to cover the regions of $x$ over which the length of the folds change, but their number remains constant. We also need to understand the merger and annihilation of folds, as in Fig.~\ref{fig-time-folds}. This takes place when the length of a given segment vanishes. In terms of the shock wave profiles, this corresponds to $h_j(x) h_{j+1}(x)\lesssim 1$. The transition is sketched in Fig.~\ref{transition}. The essential point is that two of the shocks become very weak, so the resulting geometry effectively has  two fewer shocks, in keeping with the TN.

There is an interesting point regarding nonlinearities during this transition. Classical GR nonlinearities in the collision of shocks are proportional to $h_j(x)h_{j+1}(x)$. When this product is large, the effects are strong, but the null shifts ensure that the surface $\Sigma$ passes far from the post-collision regions. However, when $h_j(x)h_{j+1}(x)$ is of order one, nonlinear effects are still important, but the shifts are small enough that $\Sigma$ can pass through the post-collision region (Fig.~\ref{transition}). We do not know the geometry in this region, and we are unable to characterize the shape of the maximal surface. Fortunately, this corresponds to a small piece of $\Sigma$, of characteristic size $\ell_{AdS}$. The reason is that as we vary $x$, the strength of the shocks changes exponentially, and the product $h_j(x)h_{j+1}(x)$ rapidly becomes much smaller than one.

\subsubsection*{The general case}
Given a general time-configuration of homogeneous shocks, the only subtlety in constructing $\Sigma$ comes from the distinction between ``through-going'' and ``switchback'' operators \cite{Stanford:2014jda}. There, as here, one can ignore through-going operators. In the localized case, shocks can also be ignorable because they are sufficiently far away that their profile at the $x$ location of interest is very weak. 

Let us therefore begin by determining which operators in a general product (\ref{what'sup}) are relevant. Let $j_1$ be the first index $j$ such that $|t_j| - |x-x_j| >t_*$. This corresponds to the first shock with appreciable strength in the frame of the maximal volume surface anchored at $t = 0$. As a recursive step, let ${j_{k+1}}$ be the least index greater than $j_k$ such that either the insertion makes $j_k$ through-going:
\begin{align}
\text{sgn}(t_{j_{k+1}}-t_{j_k}) = \text{sgn}(t_{j_{k}}-t_{j_{k-1}}) \\
|t_{j_{k+1}} - t_{j_{k-1}}| - |x - x_{j_{k+1}}|/v_B  > |t_{j_k} - t_{j_{k-1}}| - |x - x_{j_k}|/v_B 
\end{align}
or the insertion makes $j_k$ a switchback:
\begin{align}
\text{sgn}(t_{j_{k+1}}-t_{j_k}) = -\text{sgn}(t_{j_{k}}-t_{j_{k-1}}) \\
|t_{j_{k+1}} - t_{j_k}| - 2t_* - |x - x_{j_{k+1}}|/v_B - |x - x_{j_k}|/v_B  > 0.
\end{align}
The first line of the first condition ensures that the time fold is through-going at $j_k$, and the second condition ensures that the shock $h_{j_{k+1}}(x)$ is stronger than $h_{j_k}(x)$. In this case, shock $j_k$ becomes ignorable for reasons similar to those explained in \cite{Stanford:2014jda}. The first line of the second condition ensures that the fold switches back, and the second line ensures that the product of the strengths of the shocks, $h_{j_k}(x)h_{j_{k+1}}(x)$, will be at least order one. Operators failing both conditions correspond to weak shocks that only mildly affect $\Sigma$ at this transverse position $x$. After extracting this subset, we further discard all through-going shocks. To simplify notation, we re-index the remaining shocks by $j$. 

Each shock in this set now corresponds to a switchback of the time fold, so the geometry is nearly identical to the simple case considered above. The only difference is that the profiles are replaced by
\be
h_j(x) = \frac{e^{\frac{2\pi}{\beta}(\pm t_j - t_*)-\mu |x-x_j|}}{{|x-x_j|^{\frac{d-2}{2}}}}.\label{profile-different-points}
\ee
and the length of segment $(j+1)$ of the folded time axis is generalized to
\be
|t_{j+1}-t_j| - 2t_* - |x-x_{j+1}|/v_B - |x -x_j|/v_B.
\ee
Although we will not give a formal proof, one can check that this set of intervals is the solution to the minimization problem that defines the cross section of the minimal TN at location $x$, as described at the end of \S~\ref{TN}.\footnote{The $t_*$ was not present in section \ref{TN}, because the single-site ``scrambling time'' is order one for a qubit system. Its appearance in this equation is consistent with the interpretation of extra cancellation between $U(t)$ and $U(-t)$ in a large $N$ system with a single-site perturbation \cite{Stanford:2014jda}.} This implies that both the TN and the ERB geometry are fibrations of the same collection of intervals over the $x$ space, and therefore that they agree.

\section{Discussion}
Connections have been found between quantum mechanics and geometry, most notable the connection between spatial connectivity and entanglement \cite{Ryu:2006bv,VanRaamsdonk:2010pw,Maldacena:2013xja}. But entanglement may not be enough; the growth of entanglement saturates after a very short time, while geometry continues to evolve for a very long time. Evidently there is need for quantities more subtle than entanglement entropy to encode this evolution. In lattice quantum systems (in contrast to classical systems) computational complexity evolves for an exponentially long time, as does the form of the minimal tensor network describing a state. It has been conjectured that these quantities are related to the geometry behind the horizon \cite{maldacenaTN,Susskind:2014rva}.

States obtained through the action of precursor operators provide a setting to test these conjectures. In \cite{Stanford:2014jda}, spherically symmetric precursors were studied. Because of the symmetry, it was only possible to relate a single geometric quantity---ERB-volume---to a single information-theoretic quantity---complexity. By contrast, the spatially localized precursors studied in this paper provide an opportunity to relate a much wider range of local geometric properties of ERBs and TNs.

Our basic hypothesis, following \cite{maldacenaTN}, was that the structure of the  minimal tensor network, encoding the instantaneous state of the holographic boundary theory, directly reflects the Einstein geometry of the ERB. We emphasize that it is the minimal TN---the one that defines complexity---that seems to be picked out by general relativity. There are many TNs that can describe a given state. For example the naive TN on the left side of figure \ref{fig-tensor-network} generates the same state as one on the right. For a general product of precursor operators, we found a match between the ``minimalized" TN and the large-scale spatial geometry implied by general relativity.

More precisely, we found a match between the TN associated to a product of precursors in a strongly interacting lattice system, and the large-scale features of the ERB geometry characterizing an analogous product in the holographic theory. We do not mean to imply that spin systems are dual to black holes, or that we know precisely how to describe states of a continuum quantum field theory using tensor networks.\footnote{But see \cite{Susskind:2014jwa,Dowling} for a description of continuous-time complexity.} The idea is one of universality: the pattern of growth of precursor operators leads to a geometric description that is shared by a wide collection of quantum systems. 

It would be interesting to understand how wide this collection is. Strong coupling plays an important role, since precursors do not grow in free theories. A natural problem is to understand corrections to the pattern of growth at finite coupling. In gauge/gravity duality, the effects of finite gauge theory coupling translate to stringy corrections in the bulk. The role of inelastic and stringy physics in the context of precursors and shock waves is the subject of \cite{Shenker:2014cwa}.

\section*{Acknowledgments}
We are grateful to Don Marolf and Steve Shenker for discussions. D.R. is supported by the DOD through the NDSEG Program, by the Fannie and John Hertz Foundation, and is grateful for the hospitality of the Stanford Institute for Theoretical Physics. D.S. is grateful for the hospitality of the Aspen Center for Physics. This work was supported in part by National Science Foundation grants 0756174 and PHYS-1066293, and  by the U.S. Department of Energy under grant Contract Number DE-SC00012567. This publication was made possible in part through the support of a grant from the John Templeton Foundation. The opinions expressed in this publication are those of the authors and do not necessarily reflect the views of the John Templeton Foundation.

\appendix

\section{Localized shocks at finite time}\label{appendixC}

In the main text of the paper, we presented a localized shock solution proportional to $\delta(u)$. This solution is appropriate in a particular limit, where we take $t_w$ to infinity and the mass of the perturbation to zero, with $T_{uu}$ held fixed. If $t_w$ is finite, there are corrections to this solution. For example, the shock must be confined within the future lightcone of the source point. Exact solutions with this property in planar BTZ can be obtained from \cite{Horowitz:1999gf}. However, those solutions are precisely localized on the light cone, corresponding to a carefully tuned high energy insertion at the boundary. In our setting, the boundary operator should  be low energy, and the solution will not be precisely localized on the light cone. In principle, the exact linearized backreaction can be calculated using retarded propagators. Unfortunately, these functions are not known exactly for the black hole background. 

Without constructing the exact solution explicitly, we will show in this appendix that the shock profile is accurate for $x < v_B t$. If $G_N$ is small, the perturbation can be very small in this region, but it will be large compared to $G_N$.

We begin with a planar black hole metric with an arbitrary perturbation $h$:
\be
ds^2 = -A(uv)dudv + B(uv)dx^idx^i + h_{\mu\nu} dx^\mu dx^\nu.
\ee
We are interested in the response to an order one source at time $t_w$ in the past, with $t_w$ large. This solution can be constructed by applying a boost to a reference solution
\begin{align}
h_{uu}(u,v,x^i) &= \gamma^2 H_{uu}(\gamma u, \gamma^{-1}v,x^i) \\
h_{ui}(u,v,x^i) &= \gamma H_{ui}(\gamma u, \gamma^{-1}v,x^i) \\
h_{ii}(u,v,x^i) &= H_{ii}(\gamma u, \gamma^{-1}v,x^i),
\end{align}
where $\gamma = e^{\frac{2\pi}{\beta}t_w}$, and $H$ is the backreaction of a perturbation at time $t = 0$.

As we increase $\gamma$, the $h_{uu}$ and $h_{ui}$ components of the metric become larger, but the entire profile is compressed towards the $u = 0$ horizon. Physically, it is clear that the disturbance to the geometry at an order one value $u_0$ of the $u$ coordinate should remain finite as $\gamma$ grows large. That is, the effect on a late infaller of an early perturbation (on the same side) does not grow with the time separation. We will use this fact below, in the form $h_{\mu\nu}(u_0,0,x^i) \sim G_N O(1)$, where the $O(1)$ refers to $\gamma$ dependence.

Let us consider the quantity
\be
\delta v(x^i) = \int_0^{u_0} du\, \frac{h_{uu}(u,0,x^i)}{A(0)}.
\ee
For small $h$, this is the shift in the $v$ direction of a null curve near $v = 0$, traveling from $u = 0$ to $u = u_0$. We would like to show that this quantity is approximately equal to the shock wave profile $h(x)$. To do so, we consider the $u$-$u$ component of Einstein's equations. We define $E_{uu} = R_{uu} - \frac{1}{2}g_{uu}R +g_{uu}\Lambda$. To linear order in $h$, this is
\begin{align}
E_{uu} = &\frac{D-2}{2}\frac{B'}{AB}\big(-2 + v\partial_v - u\partial_u\big)h_{uu} - \frac{1}{2B}\partial_i^2h_{uu}  \notag \\&+\frac{A}{2B}\left[2\partial_u\partial_i \frac{h_{ui}}{A} - \partial_u\left(\frac{B}{A}\partial_u \frac{h_{ii}}{A}\right)\right] -\frac{v B'}{B}\frac{h_{uv}}{A}.
\end{align}
where the first term was simplified using the background equations of motion for $A$, $B$. Einstein's equations set this equal to $8\pi G_N T_{uu}$.

We will take this component of Einstein's equations, and integrate it $du$ along $v = 0$. Integrating by parts, we find
\be
(a_0 - a_1 \partial_i^2)\delta v(x^i) + a_2 \partial_i h_{ui}(u_0,0,x^i) - a_3 \partial_uh_{ii}(u_0,0,x^i) = 8\pi G_N\int_0^{u_0}du\,T_{uu}(u,0,x^i),
\ee
where the constants $a_i$ are related to $A,B,B'$ evaluated at the horizon. Since the metric components at $u_0$ are $O(1)$ at large $\gamma$, we therefore have
\be
(a_0 - a_1 \partial_i^2)\delta v(x^i)  = 8\pi G_N\int_0^{u_0}du\,T_{uu}(u,0,x^i) + G_NO(1).
\ee
This is the same equation satisfied by the shock profile. We see that it is accurate at large $\gamma$, up to a correction that is $O(1)$ in $\gamma$. The solution $\delta v(x^i)$ will therefore agree with the shock profile $G_Ne^{\frac{2\pi}{\beta}t_w - \mu |x|}$ in the region that this profile is large compared to $G_N$. That is, for $|x| < \frac{2\pi}{\beta\mu}t_w = v_Bt_w$.

\section{Localized shocks in Gauss-Bonnet}\label{GB}

\newcommand{\g}{g_{\mu\nu}}
\newcommand{\gU}{g^{\mu\nu}}
\newcommand{\Rdd}{R_{\mu\nu}}
\newcommand{\RUU}{R^{\mu\nu}}
\newcommand{\Rdddd}{R_{\mu\nu\rho\sigma}}
\newcommand{\RUUUU}{R^{\mu\nu\rho\sigma}}
\newcommand{\Rddalt}{R_{\lambda\tau}}
\newcommand{\RUUalt}{R^{\lambda\tau}}
\newcommand{\Rddddalt}{R_{\lambda\tau\rho\sigma}}
\newcommand{\RUUUUalt}{R^{\lambda\tau\rho\sigma}}
\newcommand{\RUdUd}{R^{\lambda\ \tau}_{\ \mu \ \nu}}

In Gauss-Bonnet gravity \cite{Zwiebach:1985uq} with a negative cosmological constant, the action is
\be
S=\frac{1}{16\pi G_N}\int d^{d+1}x \sqrt{-g}~\Big\{R + \frac{d(d-1)}{\ell_{AdS}^2} +
\alpha (\Rdddd \RUUUU -4\Rdd \RUU +R^2 )
\Big\},
\ee
with AdS radius $\ell_{AdS}$, and Gauss-Bonnet coefficient $\alpha$. For $d<4$, the Gauss-Bonnet term is topological. It will be convenient to rewrite this coefficient as
\be
\alpha=\frac{\lambda_{GB}~\ell_{AdS}^2}{(d-2)(d-3)},
\ee
with $\lambda_{GB}$ a dimensionless parameter.  The planar black hole solution \cite{Boulware:1985wk,Cai:2001dz,Brigante:2007nu} is
\be
ds^2=\ell_{AdS}^2\Big[-f(r)N_{\sharp}^2 dt^2+f(r)^{-1}dr^2+r^2dx^idx^i\Big], \label{gb-black-hole-metric}
\ee
with
\begin{align}
f(r)&=\frac{r^2}{2\lambda_{GB}}\Bigg[1- \sqrt{1- 4\lambda_{GB} \left(1- r^{-d}\right)}\Bigg] \label{gb-emblack} \\
N_{\sharp}^2&=\frac{1}{2}\Big(1+\sqrt{1-4\lambda_{GB}}\Big).
\end{align}
The horizon is at $r = 1$, and one can check that $f'(1) = d$, as in Einstein gravity.

Following \S~\ref{2.1}, we pass to Kruskal coordinates and assume a shock wave ansatz as the same form as \eqref{shockAnsatz}. Plugging into the GB equation of motion, with a stress tensor $T_{uu}\propto \delta(u)\delta^{d-1}(x)$, we find the condition
\be
\left(1+2\lambda_{GB}\right)\left( -\partial_i\partial_i + \mu^2 \right) h(x) \propto \delta^{d-1}(x), \label{localized-gb}
\ee
where again $\mu^2 =  \frac{d(d-1)}{2}$. The factor $(1 + 2\lambda_{GB})$ rescales the source, effectively changing the scrambling time by a small amount of order $\log (1+2\lambda_{GB})$, but the transverse dependence of $h(x)$ is unchanged. The important difference is the presence of $N_{\sharp}$ in the metric. This changes the temperature such that now $\beta = 4\pi /N_\sharp f'(1)$. The strength of the perturbation still grows with the boost factor $e^{\frac{2\pi }{\beta}t_w}$, but because the relationship between $\beta$ and $f'(1)$ has been rescaled, one finds $v_{B}(\lambda_{GB}) = N_{\sharp}v_B$.\footnote{D.S. is grateful to Juan Maldacena for pointing out a mistake in v1 of this appendix, which stated that the $e^{\frac{2\pi}{\beta}t_w}$ relationship between boosts and $t_w$ was modified. The error did not propagate elsewhere in the paper.}

\section{Maximal volume surface and decoupled surface}\label{B}
In this appendix, we will give some details about $\Sigma_{max}$, $\Sigma_{dec}$ and the relationship between them. First, let us recap the definitions:
\begin{itemize}
\item $\Sigma_{max}$ is the maximal volume codimension one surface crossing the wormhole, anchored at $t = 0$ on the two asymptotic boundaries.
\item $\Sigma_{dec}$ is the ``decoupled maximal surface.'' Specifically, it maximizes a modified volume functional, $V_{dec}$ obtained by dropping all gradients in the $x$ direction.
\end{itemize}
The coarse-grained features of these surfaces are very similar to each other, because most of the volume comes from a region where $x$ gradients are small. However, $\Sigma_{dec}$ is much easier to work with, because the defining equation is decoupled in the $x$ coordinate. Here, we will study the relationship between the surfaces in the example setting of a single shock wave.
\begin{figure}[ht]
\begin{center}
\includegraphics[scale=.7]{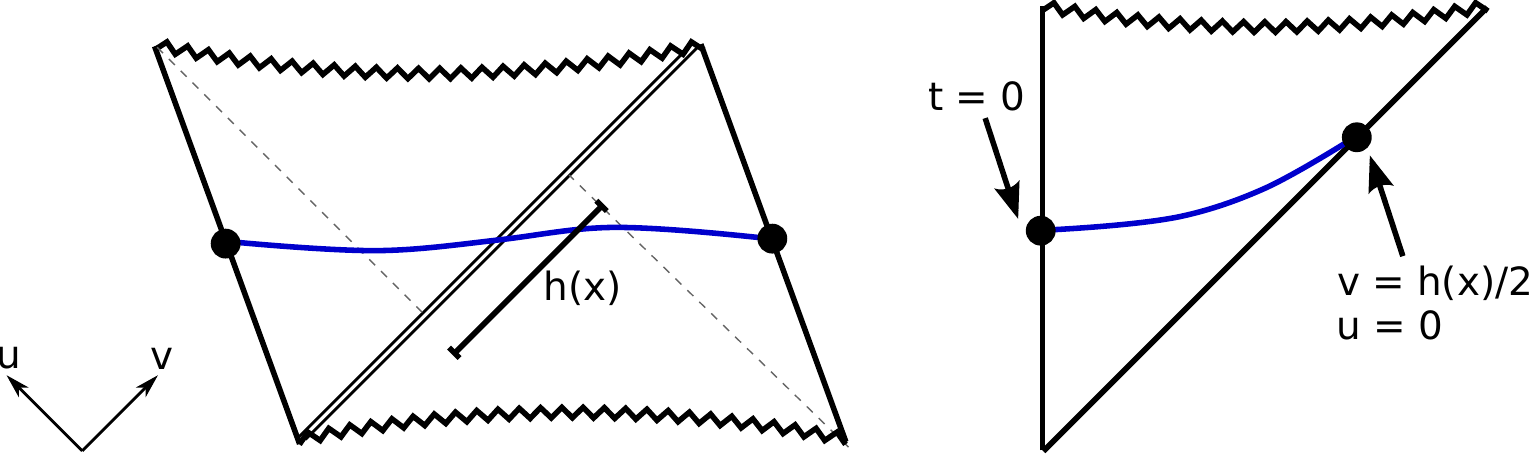}
\end{center}
\caption{{\bf Left:} a constant $x$ slice through the shock wave geometry, with the corresponding slice of the maximal volume surface shown in blue. {\bf Right:} the portion of the surface to the left of the shock is a maximal volume surface in the unperturbed black hole geometry, with the boundary conditions shown. The full surface is obtained by gluing two such pieces together.}
\label{fig-maxvol}
\end{figure}

We can use the symmetry of the shock wave geometry to reduce the problem to one in the unperturbed black hole geometry. We illustrate this in Fig.~\ref{fig-maxvol}. The left panel is a representation of a constant $x$ slice through the shock wave geometry, with a slice of the maximal volume surface shown in blue. On the right, we display only the region on one side of the shock (the other is related by a symmetry). The portion of the surface in this region is a maximal volume surface in the unperturbed black hole geometry, with boundary conditions $t = 0$ at the left boundary, and $v = h(x)/2$ at the horizon $u = 0$.

We can parametrize the surface using $v(u,x^i)$. The pullback metric is then
\be
G_{ab}dy^a dy^b = \ell_{AdS}^2\Big[ -A(uv)\partial_uv ~du^2 -A(uv)\partial_i v~du dx^i +B(uv)~dx^idx^i\Big], \label{volume-equation}
\ee
and the volume of both pieces is
\be
V=2\ell_{AdS}^d\int d^{d-1}xdu\sqrt{-AB^{d-1}\partial_u v - \frac{A^2B^{d-2}}{4}(\partial_i v)^2}. \label{max-vol-integral}
\ee
It is useful to keep in mind that $\partial_u v$ is negative, so the first term is positive. The decoupled volume $V_{dec}$ is given by dropping the second term inside the square root. 

\begin{figure}[ht]
\begin{center}
\includegraphics[scale=.25]{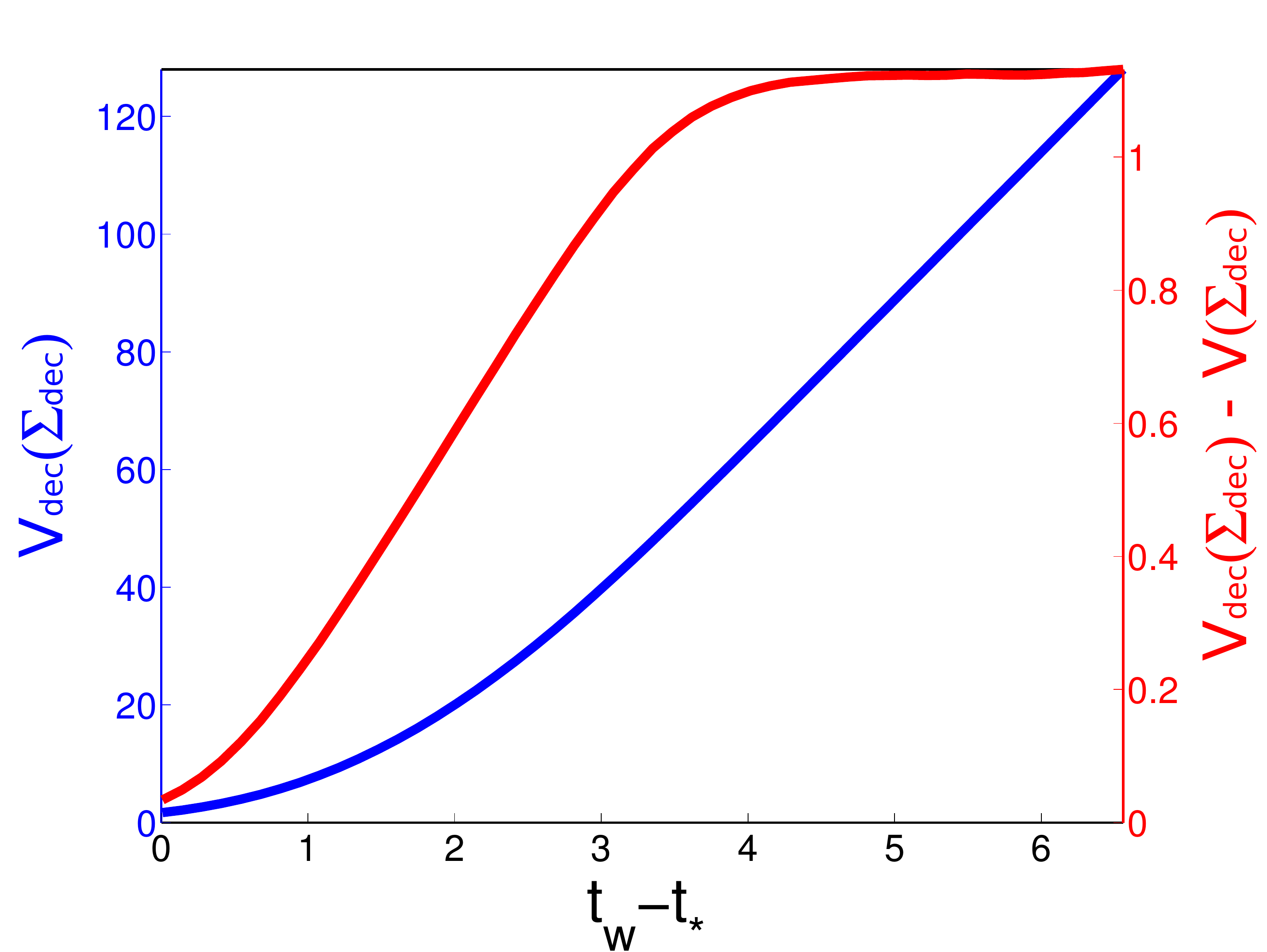}
\end{center}
\caption{The blue curve (left axis) shows the decoupled volume of the decoupled surface, $V_{dec}(\Sigma_{dec})$, as a function of $(t_w-t_*)$, for the BTZ setting of $d = 1+1$ theory on a spatial circle. Initially, the volume grows quadratically, but after a $(t_w-t_*)\sim\pi$, the size of the precursor saturates, and the volume grows linearly. The red curve (right axis) shows the gap between the upper and lower bounds in Eq.~(\ref{bounds}). The gap is quite small, and is roughly proportional to the $t_w$-derivative of the volume itself.}
\label{fig-growth-ERB}
\end{figure}

Using the surface $\Sigma_{dec}$, we can find upper and lower bounds on the volume of the maximal surface,
\be
V(\Sigma_{dec}) \le V(\Sigma_{max}) \le V_{dec}(\Sigma_{dec}).\label{bounds}
\ee
The first inequality follows from the fact that $\Sigma_{max}$ is maximal. The second follows from the fact that $V\le V_{dec}$ for any surface, and that $\Sigma_{dec}$ maximizes $V_{dec}$. In Fig.~\ref{fig-growth-ERB}, we plot $V_{dec}(\Sigma_{dec})$ and the gap in the bounds for a shock in the BTZ geometry, as a function of the strength $t_w - t_*$.
Numerically, the gap between the bounds is quite small, and appears to be proportional to the $t_w$-derivative of the volume.

We can explain this as follows. At a fixed value of $x$, the decoupled surface $v(u)$ is given by finding a maximal surface in a sptially homogeneous shock background. This problem was studied in \cite{Stanford:2014jda} using techniques from \cite{Hartman:2013qma}. For large $h(x)$, the surface tends to hug a fixed radius in the interior, $r_m$, that maximizes the function $r^{d-1}\sqrt{|f(r)|}$. In terms of $u,v$ coordinates, this special radius corresponds to a surface given by $vu = const$, independent of $x$. 

In the spatially homogeneous case, the contribution to the volume coming from the region near this surface is proportional to $\log h$. The surfaces $\Sigma_{max}$ and $\Sigma_{dec}$ will be very similar in this region, because $x$ gradients are small. To put it differently, $r = r_m$ is an attractor for both $\Sigma_{dec}$ and $\Sigma_{max}$. For large $h$, most of the surface is near this radius, and the surfaces will therefore agree at a coarse-grained level. Away from the special surface $r = r_m$, the surface $\Sigma_{max}$ and $\Sigma_{dec}$ will differ, but the regularized volume in this region is subleading at large $h$.

Integrating over $x$, the regularized decoupled volume of $\Sigma_{dec}$ will be proportional to 
\be
V_{dec}(\Sigma_{dec})\propto \int_{h(x)\ge 1}d^{d-1}x \log h(x),
\ee
while the difference will be proportional to
\be
 V_{dec}(\Sigma_{dec})-V(\Sigma_{max}) \propto \int_{h(x) \ge 1} d^{d-1}x.
\ee
It follows that the difference is subleading at large $h$, and in fact proportional to the $t_w$-derivative of the volume.

\end{document}